\documentstyle[graphicx,aps,prb,twocolumn]{revtex}

\begin{document}
\draft
%\twocolumn[\hsize\textwidth\columnwidth\hsize\csname@twocolumnfalse\endcsname
%\preprint{cond-mat/970????}
%\tightenlines

\title{Separation of Quasiparticle and Phononic Heat Currents in
       YBa$_2$Cu$_3$O$_{7-\delta }$}

\author{B.~Zeini$^1$, A.~Freimuth$^1$, B.~B\"{u}chner$^1$, R.~Gross$^1$,
A.P.~Kampf$^2$, M.~Kl\"{a}ser$^3$, and G.~M\"{u}ller-Vogt$^3$}

\address{$^1$ II. Physikalisches Institut, Universit\"{a}t zu K\"{o}ln,
              50937 K\"{o}ln, Germany \\
         $^2$ Theoretische Physik III, Universit\"{a}t Augsburg, 86135 Augsburg,
              Germany \\
         $^3$ Kristall- und Materiallabor, Universit\"{a}t Karlsruhe,
              76128 Karlsruhe, Germany}
\date{Received 9 February 1998}
\maketitle

\begin{abstract}
\noindent Measurements of the transverse ($k_{xy}$) and longitudinal ($k_{xx}$)
thermal conductivity in high magnetic fields are used to separate the
quasiparticle thermal conductivity ($k_{xx}^{el}$) of the CuO$_2-\rm
planes$ from the phononic thermal conductivity in $\rm
YBa_2Cu_3O_{7-\delta }$. $k_{xx}^{el}$ is found to display a
pronounced maximum below $T_c$. Our data analysis reveals distinct
transport ($\tau $) and Hall ($\tau_H$) relaxation times below $T_c$:
Whereas $\tau$ is strongly enhanced, $\tau_H$ follows the same
temperature dependence as above $T_c$.
\end{abstract}

%\maketitle
%\date{\today}
%]
%\narrowtext

The study of heat transport in the superconducting state is well known
to provide valueable information on the quasiparticle (QP) excitations
and their dynamics. Compared to other probes of the QP-dynamics such
as the microwave conductivity thermal transport has the advantage of
probing only the QP-response, since the superfluid does not carry
heat. On the other hand, a major complication in the analysis of the
thermal conductivity is often a substantial phononic contribution
$k_{xx}^{ph}$ to the heat current. Such a situation is realized in the
high-$T_c$ superconductors (HTSC), where the heat current above $T_c$
is known to be dominated by $k_{xx}^{ph}$~\cite{Uher}. Accordingly,
the interpretation of experimental data is ambiguous: For example, the
maximum in the temperature dependence of the thermal conductivity in
the superconducting state of $\rm YBa_2Cu_3O_{7-\delta}$ (YBCO) has
been attributed to both a maximum of the electronic as well as the
phononic contribution~\cite{Uher,Yu,Peacor,Cohn}. A clear separation
of the phonon and QP heat currents is difficult and up to now an
unsolved problem.

It has been pointed out in Refs.~\cite{Freimuth1,Krishana} that the
transverse thermal conductivity $k_{xy}$ -- the thermal analogue of
the Hall effect -- is free of phonons, i.e. $k_{xy}$ is purely
electronic. $k_{xy}$ is related to the electronic thermal conductivity
$k_{xx}^{el}$ according to $k_{xy}
= k_{xx}^{el}\tan\alpha_R$ where $\alpha_R$ is the thermal Hall angle.
From the Wiedemann-Franz law~\cite{WFG} one expects that $\alpha_R$ is
equal to the electrical Hall angle $\alpha_H$ as obtained from the
electrical ($\sigma_{xx}$) and Hall ($\sigma_{xy}$) conductivities via
$\tan \alpha_H = \sigma_{xy}/\sigma_{xx}$. In conventional metals
$\tan\alpha_H=\omega_c \tau$ where $\omega_c=eB/m$ is the cyclotron
frequency and $\tau$ is the usual transport relaxation time. In
contrast, in the normal state of the HTSC $\tan \alpha_H$ is highly
anomalous and has a temperature dependence which is {\em distinctly}
different from $\tau(T)$~\cite{Iye,Chien,Anderson}. Up to now no data
on the QP Hall angle are available below $T_c$, but one may suspect
that its behavior is anomalous as well. Therefore, in order to
calculate $k_{xx}^{el}$ from $k_{xy}$ it is necessary to treat
$\tan\alpha_R=\omega_c\tau_R$ as an additional independent parameter
which must be determined experimentally.

In this letter we determine both $k_{xx}^{el}$ and $\tan \alpha_R$ below $T_c$
in a single crystal of YBCO from combined measurements of $k_{xy}$ and $k_{xx}$
in high magnetic fields. Our main results are: (1) The electronic thermal
conductivity of the $\rm CuO_2$-planes shows a pronounced maximum below $T_c$
which is strongly suppressed by a magnetic field. (2) $\tan\alpha_R$, as
extracted from the thermal transport data below $T_c$, displays the same
temperature and magnetic field dependence as $\tan\alpha_H$ obtained from
electrical transport data above $T_c$, and it passes smoothly through $T_c$.
This shows that $\tau_R \simeq \tau_H$ and, remarkably, that $\tau_H$ and
$\tau $ behave differently also below $T_c$!

The thermal conductivity tensor $\underline{k}$ is the sum of an
electronic and a phononic part
$\underline{k}=\underline{k}^{el}+\underline{k}^{ph}$. It is defined
via the heat current density ${\bf j}_h=-\underline{k}\nabla
T$~\cite{Vormo}. We assume that $\underline{k}^{ph}$ remains diagonal
even for ${\bf B}\neq 0$, i.e. $k_{xy}^{ph}=0$. In this case the
transverse components of $\underline{k}$ are purely electronic. The
transverse thermal conductivity, also called the Righi-Leduc-effect,
is measured as follows: In a magnetic field ${\bf B}=(0,0,B)$ a
temperature gradient $\nabla_x T$ is applied in the $x$-direction.
Under the condition $j_{h,y}=0$ a transverse temperature gradient
$\nabla_y T$ is found in the $y$-direction. Using the Onsager
relations we find in this situation
\begin{equation}
\label{kxy}
j_{h,y} = k_{xy} \nabla_x T - k_{xx} \nabla_y T = 0 \,
\end{equation}
with $k_{xx}=k_{yy}$ for twinned crystals without in-plane anisotropy.
$k_{xy}$ can therefore be determined experimentally by measuring $\nabla_x T$,
$\nabla_y T$, and the {\em total} longitudinal thermal conductivity $k_{xx}$.

Our measurements were carried out at constant temperatures with the magnetic
field applied perpendicular to the $\rm CuO_2$-planes. Typically temperature
gradients $\nabla_x T$ of order 0.5 K/mm were applied using a small manganin
heater mounted on top of the samples. The resulting transverse temperature
gradients $\nabla_y T$ of order 10$^{-3}$ K/mm in magnetic fields up to 14 T
were measured with AuFe-Chromel thermocouples calibrated in the same field
range~\cite{calibration}. To eliminate offset voltages due to misalignment
of the thermocouple we have
measured for both field directions $\pm {\bf B}$ in order to determine the
Righi-Leduc component of $\nabla_y T$ which must be antisymmetric with respect
to field reversal. We have measured in 2 different modes: Either ${\bf B}$ was
reversed at fixed temperature or we have heated the sample to temperatures
above $T_c$ before the field was reversed. Because of vortex pinning effects
this latter mode was used for all low temperature measurements. We note that
sweeping the magnetic field at fixed temperatures results in a strong hysteresis
of $k_{xx}$ in almost the entire temperature range below $T_c$, similar as
reported in Ref.~\cite{Aubin}; $k_{xy}$, which must be extracted from the
asymmetry of the field dependence, can therefore not be determined by
sweeping the magnetic field. We have tested our method by measurements
on an insulator ($k_{xy} = 0$) and on simple
metals~\cite{Zeini}. Details of our experimental setup will be described
elsewhere. The results presented here have been obtained on a high quality
twinned single crystal of YBCO with dimensions 1.9 mm $\times $ 2 mm $\times $ 0.38 mm
and with a superconducting transition at $T_c \simeq 90.5$K.

Representative experimental results are shown in Fig.~1. $k_{xx}$ has a
pronounced maximum at $T_{max} < T_c$ which is strongly suppressed by the
applied magnetic field. The absolute value of $k_{xx}$ ($\approx $10 W/Km at
$T_c$), the relative upturn of $k_{xx}$ in zero field as characterized by
the ratio $k_{xx}(T_{max})/k_{xx}(T_c) \approx 1.6$ for our sample, as well as
the sensitivity of the maximum to magnetic fields are consistent with previous
results~\cite{Uher,Yu,Peacor}. The overall temperature dependence of $k_{xy}$
is similar to that of $k_{xx}$ but the maximum of $k_{xy}$ occurs at higher
temperatures and the relative change below $T_c$ is larger in comparison to
$k_{xx}$. The absolute magnitude of $k_{xy}$ is comparable to that reported
previously in Ref.~\cite{Krishana}.

\begin{figure}[b]
  \begin{center}
    \setlength{\unitlength}{1cm}
    \includegraphics[width= 0.87\columnwidth, clip]{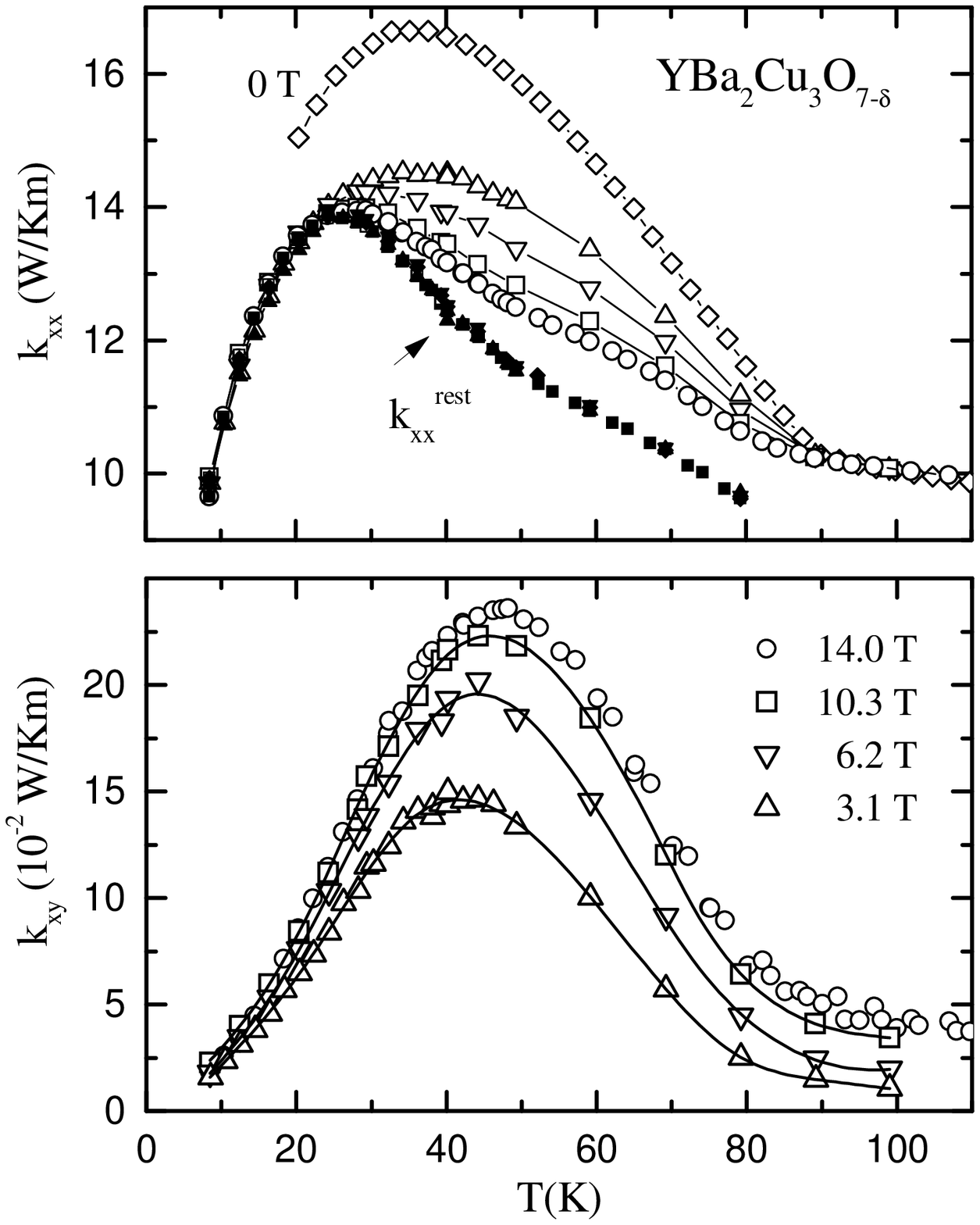}
   \end{center}
\caption{Open symbols: $k_{xx}$ (upper panel) and $k_{xy}$ (lower panel) of
         $\rm YBa_2Cu_3O_{7-\delta }$  as a function of temperature $T$ for
         various fixed magnetic fields as indicated in the figure. Full
         symbols: $k_{xx}^{rest}$ as obtained from our data analysis (see
         text).}
\label{fig1}
\end{figure}

For our data analysis we assume that in YBCO 3 channels of heat conduction are
present:
\begin{equation}
\label{ksum}
k_{xx} = k_{xx}^{el} + k_{xx}^{ch} + k_{xx}^{ph}  = k_{xx}^{el} +
k_{xx}^{rest}  \, .
\end{equation}
Here $k_{xx}^{el}$ is the electronic contribution from the $\rm CuO_2$-planes
($\rm CuO_2$-bilayers in YBCO) and $k_{xx}^{ph}$ is that of the phonons.
$k_{xx}^{ch}$ describes a possible contribution from the $\rm CuO$-chains
which are present in YBCO along the $b$-direction of the orthorhombic
structure. These chains are metallic for optimally doped samples and lead to a
rather strong $a$-$b$-anisotropy in untwinned crystals
($\sigma_{bb}/\sigma_{aa} \simeq 2$)~\cite{Iye}. In a twinned crystal they
should contribute to the electrical and the heat conduction on average;
$a$-$b$-anisotropy is of course absent. In the subsequent data analysis
$k_{xx}^{ch}$ must be treated differently from $k_{xx}^{el}$ since the
CuO-chains as a quasi-1-dimensional channel for charge and heat transport
should neither contribute to the transverse transport coefficients
$\sigma_{xy}$ and $k_{xy}$ nor to the magnetic field dependence of the
longitudinal ones.

\begin{figure}[b]
  \begin{center}
    \setlength{\unitlength}{1cm}
    \includegraphics[width= 0.9\columnwidth, clip]{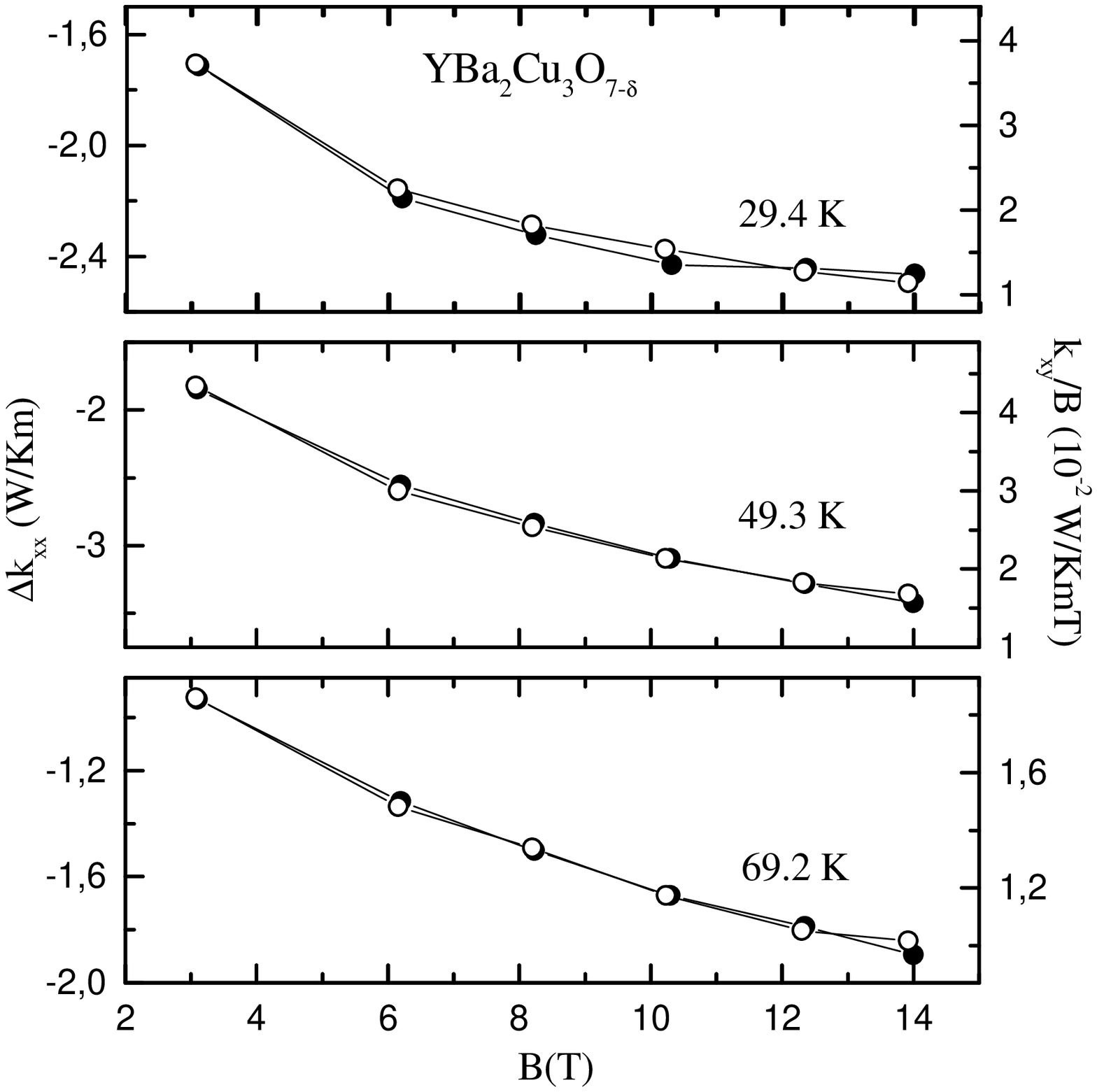}
   \end{center}
\caption{$\Delta k_{xx} = k_{xx}(B) - k_{xx}(B=0)$ ($\bullet $) and
         $k_{xy}/B$ ($\circ $) versus magnetic field $B$ at fixed
         temperatures given in the figure.}
\label{fig2}
\end{figure}

We compare in Fig.~\ref{fig2} $\Delta k_{xx} = k_{xx}(B) - k_{xx}(0)$ and
$k_{xy}/B$. Notably, these quantities have the same magnetic field dependence,
i.e.
\begin{equation}
\label{result}
\frac{\partial }{\partial B} \left( \frac{k_{xy}}{B} \right) \propto
\frac{\partial k_{xx}}{\partial B} \, .
\end{equation}
This observation provides the key to our data analysis. We define
\begin{equation}
\label{taur}
\tan \alpha_R = \omega_c \tau_R = \frac{k_{xy}}{k_{xx}^{el}} \, ,
\end{equation}
where $\tau_R(B,T)$ is a ''relaxation time'' introduced to parametrize the
field and temperature dependence of $\tan \alpha_R$. Using Eqs.~(\ref{ksum})
and (\ref{taur}) we find
\begin{eqnarray}
\label{affe}
\frac{m}{e}\frac{\partial }{\partial B}\left(\frac{k_{xy}}{B} \right)=
\tau_R\frac{\partial k_{xx}}{\partial B}\\
+ \left[ k_{xx}^{el} \frac{\partial \tau_R}{\partial B} - \tau_R
\frac{\partial
  k_{xx}^{ph}}{\partial B} - \tau_R \frac{\partial k_{xx}^{ch}}{\partial B}
\right] \,. \nonumber
\end{eqnarray}
Apparently, our experimental results suggest that the term in brackets is
zero. Since the 3 terms in brackets refer to 3 distinct channels of heat
conduction this requires that $\tau_R$, $k_{xx}^{ph}$, and $k_{xx}^{ch}$ are
separately field independent. Note that this is certainly reasonable in view of
the origin of these contributions to the heat current.

Assuming thus that the 3 terms in brackets vanish $\tau_R$ can be calculacted from our
data by comparing $k_{xx}$ and $k_{xy}/B$ at different magnetic fields
according to $e\tau_R/m =\Delta (k_{xy}/B)/\Delta k_{xx}$ (see
Eq.~(\ref{affe})). The result is shown in Fig.~3. Note that the values
obtained for different magnetic fields coincide within the experimental
accuracy consistent with the anticipated field independence of $\tau_R$.

As a check of our result for $\tau_R$ we have also determined $e\tau_H/m =
\sigma_{xy}/B\sigma_{xx}$ for the same sample from measurements of
$\sigma_{xy}$ and $\sigma_{xx}$ in the normal state~\cite{Zeini}. These data
as well as their extrapolation~\cite{fitformula} to temperatures below $T_c$
are also shown in Fig.~3. The extrapolated values for $\tau_H^{-1}$ look very
similar to $\tau_R^{-1}$ regarding the temperature dependence (see inset of Fig.~3),
but they appear to be systematically larger by roughly a factor
2. However, note that $\tau_R$ as extracted from the thermal transport data is
clearly unaffected by the CuO-chains and that this is also true for
$\sigma_{xy}$. In contrast, $\sigma_{xx}$ has a contribution from the
CuO-chains, i.e.
$\sigma_{xx} = \sigma_{xx}^{pl} + \langle \sigma_{xx}^{ch} \rangle$,
where $\sigma_{xx}^{pl}$ is the electrical conductivity of the $\rm
CuO_2$-planes and $\langle \sigma_{xx}^{ch} \rangle$ is an average
of the chain contribution appropriate for a twinned crystal. With
$\sigma_{xx}^{pl} \approx \langle \sigma_{xx}^{ch} \rangle$~\cite{Iye} we
conclude that $e \tau_H/m = \sigma_{xy}/B\sigma_{xx}$ is underestimated by
a factor 2. Correcting the normal state data for this factor we find excellent
agreement between $\tau_H$ and $\tau_R$, i.e. our data tell
$\tau_R\simeq\tau_H$. This strongly supports the procedure of our data
analysis.

Once $\tau_R$ is known the remainder of our analysis is
straightforward: $k_{xx}^{el}(B\neq 0)$ follows from Eq.~(\ref{taur})
using the data for $k_{xy}(B)$, and $k_{xx}^{rest}$ is obtained
subsequently from Eq.~(\ref{ksum}) for each field strength. As a test
for internal consistency we have verified that $k_{xx}^{rest}$ is
indeed field independent. Finally, $k_{xx}^{el}(B=0)$ follows from
Eq.~(\ref{ksum}) using $k_{xx}^{rest}$ and the zero field data for
$k_{xx}$. We have also determined $k_{xx}^{el}(B=0)$ directly from
Eq.~(\ref{taur}) using an extrapolation of $B/k_{xy}$ to $B=0$ in good
agreement with the results obtained from using $k_{xx}^{rest}$ and
Eq.~(\ref{ksum}).

In Fig.~4 we show $k_{xx}^{el}(B)$ as obtained from our data analysis.
$k_{xx}^{el}$ represents the electronic thermal conductivity of the
$\rm CuO_2$-planes in YBCO. Our results thus confirm explicitly that
$k_{xx}^{el}$ is strongly enhanced below $T_c$.
$k_{xx}^{el}\propto Tn_{QP}\tau$ implies that $\tau $ is strongly enhanced
below $T_c$ overcompensating the decrease of the QP number density $n_{QP}(T)$
with decreasing temperature. This confirms the results obtained for the QP
relaxation time from the microwave conductivity~\cite{Bonn}.

\begin{figure}[t]
  \begin{center}
    \setlength{\unitlength}{1cm}
    \includegraphics[width= 0.65\columnwidth, clip]{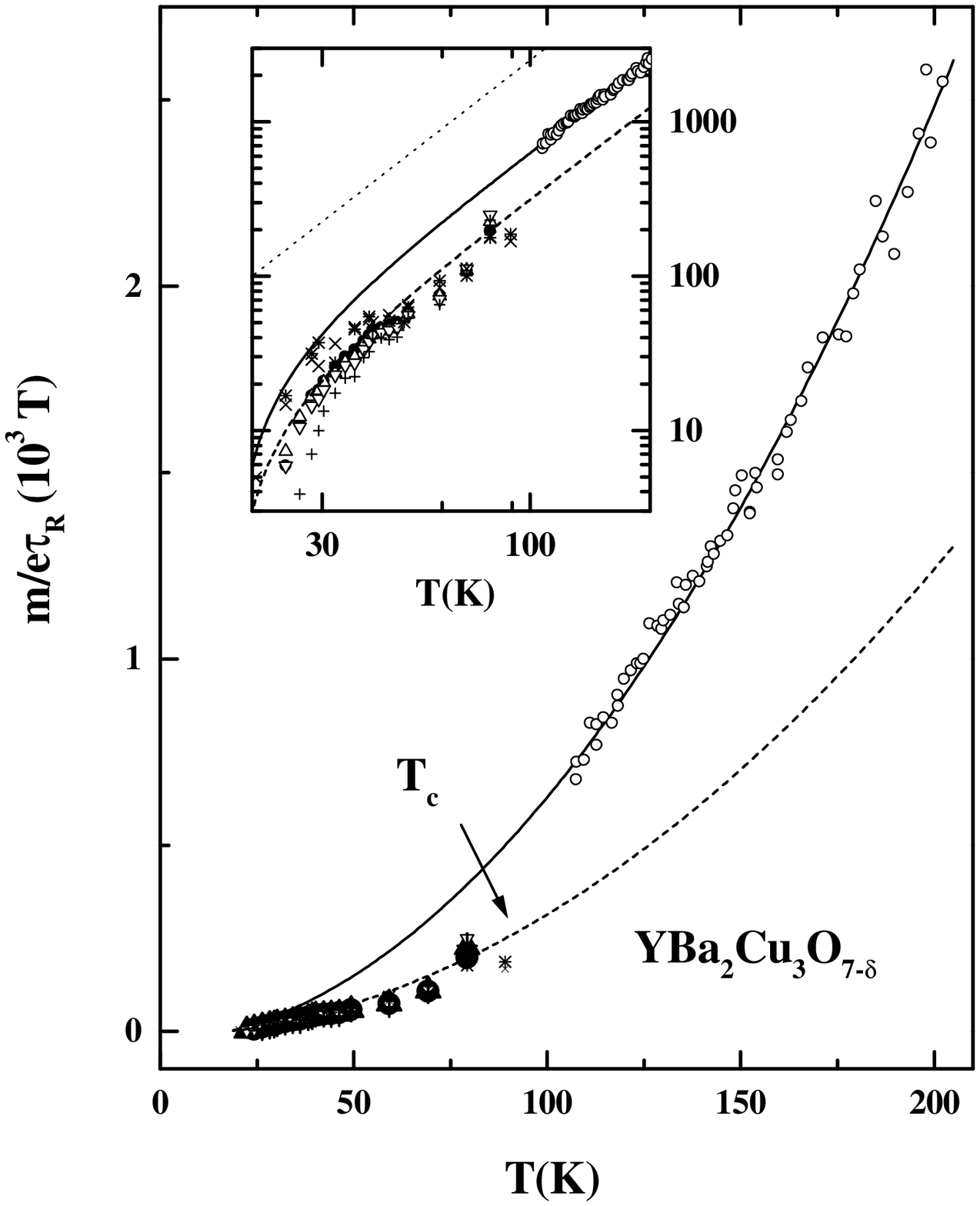}
   \end{center}
\caption{$T<T_c$: $m/e\tau_R$ vs. temperature $T$ obtained from
         $e\tau_R/m=\Delta (k_{xy}/B) /\Delta k_{xx}$, where
         $\Delta(k_{xy}/B)=k_{xy}(B_1)/B_1-k_{xy}(B_2)/B_2$ and
         $\Delta k_{xx} = k_{xx}(B_1) - k_{xx}(B_2)$. Different symbols
         correspond to different values of $B_1$ and $B_2$.
         $T>T_c$: $m/e \tau_H$ obtained from $\sigma_{xy}$ and
         $\sigma_{xx}$. Solid line: extrapolated normal state data (see text).
         Dashed line: extrapolated normal state data divided by a factor 2.
         Inset: The same data on a double logarithmic scale. The dotted line corresponds
         to a $T^2$ temperature dependence.}
\label{fig3}
\end{figure}

We also find that $k_{xx}^{el}$ is very sensitive to magnetic fields contrary
to what is observed in the normal state, where the total thermal conductivity
and thus $k_{xx}^{el}$ is field independent. It is straightforward to attribute
the field dependence below $T_c$ to an additional scattering
mechanism~\cite{nqp} characteristic for the superconducting state such as
scattering of QPs on vortices~\cite{Krishana,Salamon,Cleary}. Assuming the
corresponding scattering rate $\tau_v^{-1}$ to be proportional to the number
of vortices $n_v\propto B$ one expects for the total scattering rate that
$\tau^{-1}=\tau_{in}^{-1}+\tau_v^{-1}=\tau_{in}^{-1}+\alpha B$. Here,
$\tau_{in}$ includes the same scattering processes as in the normal state,
i.e. it has in general an elastic defect and an inelastic contribution; the
latter collapses below $T_c$. Using
$k_{xy}=k_{xx}^{el}\omega_c\tau_R=L(T)T \sigma_{xx} \omega_c \tau_R$ where $L$
is the Lorentz-number and $\sigma_{xx} = n_{QP}(T)  e^2 \tau/m$ we find
\begin{equation}
\frac{B}{k_{xy}} = C(T) \tau^{-1} = C(T) \left( \tau_{in}^{-1} + \alpha B
 \right)  \, ,
\end{equation}
where $C(T)=m^2/(LT n_{QP}e^2\tau_R)$ depends only on temperature. Thus a plot
of $B/k_{xy}$ vs. $B$ should yield a straight line. This is indeed the case as
shown in Fig.~4. We note that by assuming $\alpha$ to be temperature independent
$\tau_{in}(T)$ can be extracted from the slope and the intersection of the
$B/k_{xy}$ vs. $B$ curves (to within the constant factor $\alpha$). We find that
$\tau_{in}$ increases strongly below $T_c$ ~\cite{Zeini}.

\begin{figure}[t]
  \begin{center}
    \setlength{\unitlength}{1cm}
    \includegraphics[width= 0.85\columnwidth, clip]{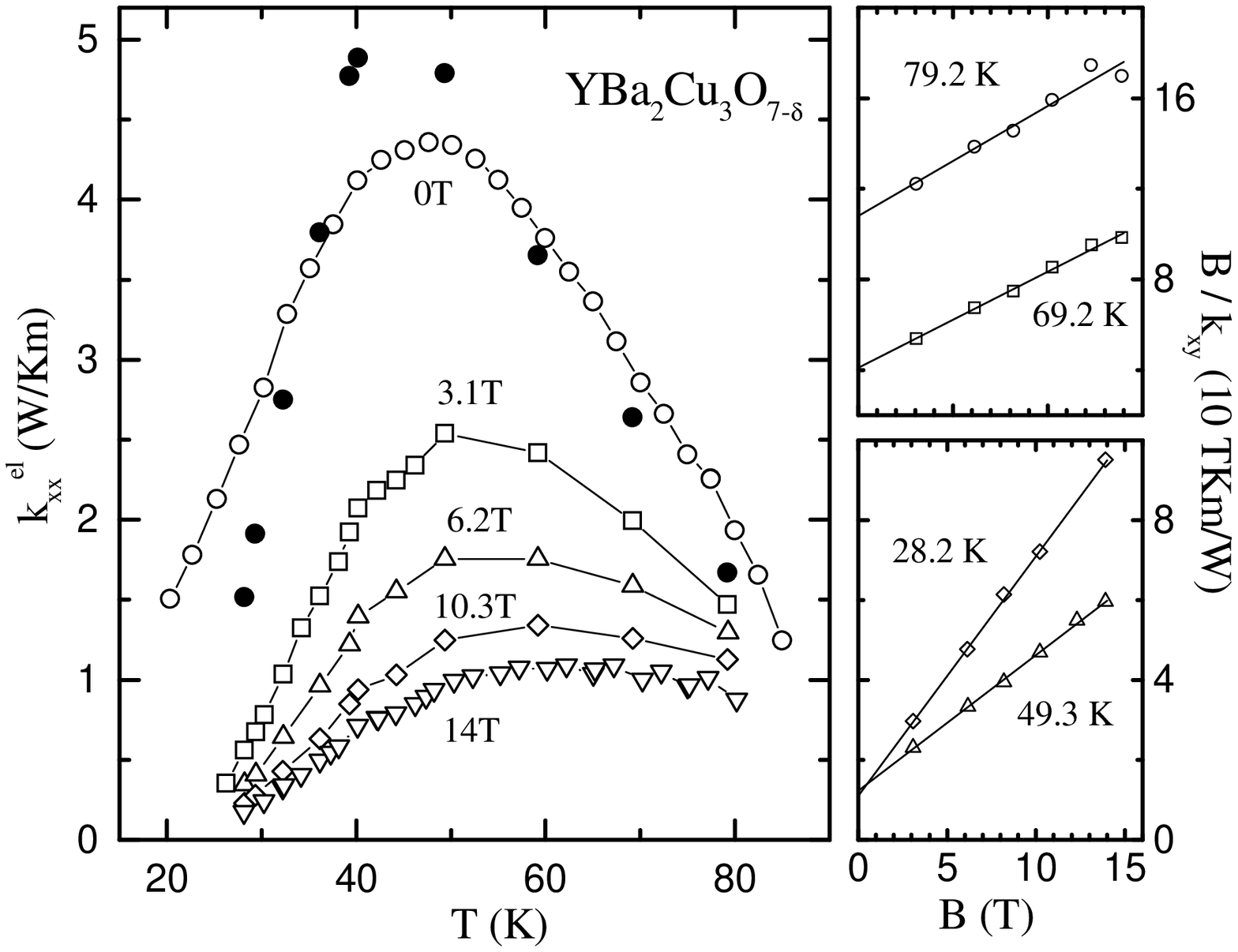}
   \end{center}
\caption{Left panel: Electronic thermal conductivity $k_{xx}^{el}(B)$ of the
         $\rm CuO_2$-planes as a function of temperature for various magnetic
         fields as indicated in the figure. The zero field data have been
         obtained from $k_{xx}^{el}(B=0)=k_{xx}(B=0)-k_{xx}^{rest}$ ($\circ$)
         as well as from an extrapolation of $B/k_{xy}$ to $B=0$ ($\bullet $).
         Right panel: $B/k_{xy}$ as a function of magnetic field $B$ at various
         fixed temperatures given in the figure.}
\label{fig4}
\end{figure}

$k_{xx}^{rest}$ as obtained from our data analysis is shown in Fig.~1.
Remarkably, $k_{xx}^{rest}$ shows a pronounced maximum below $T_c$,
too. This maximum may be due to the phononic contribution~\cite{Uher}.
However, it may also arise from the chain contribution. The latter has
previously been determined experimentally from the $a$-$b$-anisotropy
of $k_{xx}$ in detwinned single crystals of YBCO~\cite{Yu,Gagnon}.
$k^{ch}_{xx}$ shows a pronounced maximum below $T_c$ with an overall
temperature dependence similar to that found here for
$k_{xx}^{rest}$~\cite{Gagnon}. Furthermore, the field independence of
$k_{xx}^{rest}$ implies that the phononic contribution $k_{xx}^{ph}$
is independent of $B$. Such a conclusion has recently been drawn also
on the basis of low temperature results for the thermal conductivity
in Bi-based HTSC~\cite{Ong}.

Finally, our data show clearly that $\tau_R \simeq \tau_H$ and that --
as in the normal state -- $\tau_H$ and $\tau $ behave differently also
below $T_c$. In particular, whereas $\tau $ is strongly enhanced below
$T_c$, $\tau_H$ is unaffected by the superconducting transition and
shows the same temperature dependence as above $T_c$. This finding
should provide important information for the theoretical understanding
of transport phenomena in the cuprates.

In summary, we have presented a separation of the QP and phononic contributions
to the thermal conductivity below $T_c$ in YBCO based on measurements of the
longitudinal and transverse thermal conductivity in high magnetic fields. Our
data analysis shows explicitly that the QP contribution to $k_{xx}$ is strongly
enhanced below $T_c$ and that it is the QP contribution to the heat current
which is responsible for the magnetic field dependence of $k_{xx}$. We find
that -- as in the normal state -- 2 relaxation times must be distinguished also
below $T_c$: Whereas the QP relaxation time $\tau$ is strongly enhanced below
$T_c$ and magnetic field dependent, the Hall relaxation time $\tau_H$ remains
independent of $B$ below $T_c$ and has the same temperature dependence as
above $T_c$.

\begin{acknowledgements}
We are particularly grateful for stimulating discussions with W.
Brenig, Ch. Bruder,  M. Galffy, P.J. Hirschfeld, T. Kopp, D. Rainer,
S. Uhlenbruck, and P. W\"{o}lfle. This work was supported by the
Deutsche Forschungsgemeinschaft through SFB 341.
\end{acknowledgements}

%\end{multicols}

\begin{references}
\bibitem{WFG}         The generalized Wiedemann-Franz law relates the
                      electrical and thermal conductivity tensors,
                      $\underline{\sigma}$ and $\underline{k}$, respectively,
                      according to $\underline{k} = L T \underline{\sigma}$.
                      $L$ is the Lorentz number.
\bibitem{Vormo}       The heat current due to vortex motion is negligibly small [14].
\bibitem{fitformula}  We have fitted the resistivity $\rho $ and the inverse
                      Hall coefficient $R_H^{-1}$ to $a+bT$ and to
                      $\tilde{a}+\tilde{b}T$, respectively, where $a$, $b$,
                      $\tilde{a}$, and $\tilde{b}$ are constants.
                      $e\tau_R/m$ was then obtained from
                      $e\tau_H/m=\sigma_{xy}/B\sigma_{xx}
                      \simeq R_H/\rho$.
\end{references}
\end{document}